\documentstyle[12pt]{article}

\renewcommand{\a}{\alpha}
\renewcommand{\b}{\beta}
\renewcommand{\c}{\gamma}
\renewcommand{\d}{\delta}

\newcommand{\la}{\lambda}
\newcommand{\k}{\kappa}

\newcommand{\shalf}{\frac{1}{2}}
\newcommand{\pa}{\partial}
\begin{document}

\topmargin 0pt
\oddsidemargin 5mm

\newcommand{\NP}[1]{Nucl.\ Phys.\ {\bf #1}}
\newcommand{\AP}[1]{Ann.\ Phys.\ {\bf #1}}
\newcommand{\PL}[1]{Phys.\ Lett.\ {\bf #1}}
\newcommand{\NC}[1]{Nuovo Cimento {\bf #1}}
\newcommand{\JMP}[1]{Jour.\ Math.\ Phys.\ {\bf #1}}
\newcommand{\PR}[1]{Phys.\ Rev.\ {\bf #1}}
\newcommand{\PRL}[1]{Phys.\ Rev.\ Lett.\ {\bf #1}}
\newcommand{\JL}[1]{JETP.Lett.\ {\bf #1}}
\newcommand{\JETP}[1]{\ Jour.\ Eksp.\ Teor.\ Phys.\ {\bf #1}}
\newcommand{\YP}[1]{Yad.\ Phys.\ {\bf #1}}
\renewcommand{\thefootnote}{\fnsymbol{footnote}}

\begin{titlepage}
\setcounter{page}{0}
\rightline{Preprint YERPHY-1399(10)-93}

\vspace{2cm}
\begin{center}
{\Large Canonical quantization of the D=2n
dimensional relativistic spinning particle
 with anomalous magnetic moment
  in the external electromagnetic
 field}
\vspace{1cm}

{\large Grigoryan G.V.,Grigoryan R.P.} \\
\vspace{1cm}
{\em Yerevan Physics Institute, Republic of Armenia}\\
\end{center}

\vspace{5mm}
\centerline{{\bf{Abstract}}}
The pseudoclassical hamiltonian and action of the
$D=2n$ dimensional Dirac particle with anomalous
magnetic moment interacting with the external
electromagnetic field is found. The
Bargmann-Michel-Telegdi equation of motion for
the Pauli-Lubanski vector is deduced.
The canonical quantization of $D=2n$ dimensional
Dirac spinning particle with anomalous magnetic moment
in the external electromagnetic
field is carried out in the gauge which allows
to describe simultaneously particles and antiparticles
 (massive and massless) already at the classical level.
Pseudoclassical Foldy-Wouthuysen transformation is used
to obtain canonical (Newton-Wigner) coordinates and in
terms of this variables the theory is quantized.
 The connection of this quantization with the
deGroot and Suttorp's description of Dirac particle
  with anomalous magnetic moment
 in the external electromagnetic field is discussed.

\vfill
\centerline{\large Yerevan Physics Institute}
\centerline{\large Yerevan 1993}

\end{titlepage}
\newpage
\renewcommand{\thefootnote}{\arabic{footnote}}
\setcounter{footnote}{0}

\section{Introduction}
\indent

It is well known [1-4]
, that pseudoclassical description of the spinning
 particle is a theory with constraints which
 is supersymmetry and reparametrization invariant.
 Dirac quantization of this theory, when the first
 class constraints turn into a equations for wave fuctions,
leads to a covariant relativistic wave equation for a
$D=4$ dimensional spin $\shalf$ particle without anomalous
magnetic moment (AMM). The quantization of the
 particle with AMM was considered in \cite{BCL}, however the
 authors of that article came to a conclusion, that
 the pseudoclassical particle with AMM
  can not be consistently quantized.
 The reason for such a conclusion was the wrong
 supersymmetry generator, which after quantization
 didn't lead to a wave equation of spin $\shalf$
 particle with AMM. In this paper we show that if
  one starts with the right
 supersymmetry generator, then one can consistently
 quantize the theory.

The paper is organized as follows. In sect.2 we derive
the hamiltonian and the action (lagrangian)
of $D=2n$ dimensional
 spin $\shalf$ particle with anomalous magnetic
moment in the external electromagnetic field.
In sect.3 the Bargmann-Michel-Telegdi (BMT) equation for
the Pauli-Lubanski vector in $D=2n$ is found from the
hamiltonian obtained in sect.2. Note, that for $D=4$
the BMT equation was found in \cite{BM} from the action
which didn't contain  some of
the  terms in the action of section 2 proportional to
AMM, which however contribute
only to quantum corrections to the BMT equation.
In sect.4 we fix all gauge degrees of freedom and carry
out the canonical quantization in the same lines as in
\cite{GG3},\cite{GG1}. We use pseudoclassical Foldy -Wouthuysen
transformation \cite{BCLFW} to find out
the canonical variables,
 in terms of which the quantization of the theory will
  be performed. The quantized theory in $D=4$
  dimensions is an extension to an arbitrary
  electromagnetic field of the deGroot and Suttorp
  generalization \cite{SG} of the Blount picture
   to describe the particle with AMM.

\section{The Action of spin $\shalf$ particle with
 anomalous magnetic moment}
\indent

To find out the action for $D=2n$ dimensional
spinning particle with AMM we will suppose
that primary
constraints for the Grassmann variables $\xi_\mu,
\mu=1,\ldots,D-1$ and $\xi_{D+1}$ describing
the spin degrees of freedom are the same as in
the case of the particle without AMM. Namely
\begin{equation}
\label{ConstraintsA}
\Phi_\mu=\pi_\mu-\frac{i}{2}\xi_\mu,\quad
\Phi_D=\pi_{D+1}-\frac{i}{2}\xi_{D+1},
\end{equation}
where $\pi_\mu, \pi_{D+1}$ are the momenta
conjugate to $\xi_\mu$ and $\xi_{D+1}$
 correspondingly. The Dirac brackets
 ${\{\ldots,\ldots\}}^*$ of the variables of the
 theory for this set of constraints
  are given by the relations
\begin{equation}
\label{PrelimDB}
{\{x_\mu,{\cal P}_\nu\}}^* = g_{\mu\nu},\quad
\{{\xi_\mu,\xi_\nu\}}^* =
ig_{\mu\nu}, \quad {\{\xi_{D+1},\xi_{D+1}\}}^* = -i,
\quad {\{{\cal P}_\mu,{\cal P}_\nu\}}^* = gF_{\mu\nu}.
\end{equation}
 (all other brackets vanish). Here the variables
  $x^\mu$ are the coordinates of the
particle, ${\cal P}_\mu=P_\mu-gA_\mu$, $P_\mu$ are
 the momenta conjugate to $x_\mu$, $g$ is the charge of
  the  particle, $A^\mu$ is  the vector-potential of the
 electromagnetic field, $F_{\mu\nu} = \pa_\mu A_\nu -
 \pa_\nu A_\mu$.

The quantization using Dirac brackets (\ref{PrelimDB})
will bring to the Dirac equation for a particle with AMM
if we take instead of the constraint $\Phi_{D+1}=
 {\cal P}_\mu\xi^\mu-m\xi_{D+1}$ \cite{BCL} the
constraint
\begin{equation}
\label{ConstraintD1}
\Phi_{D+1}={\cal
P}_\mu\xi^\mu-m\xi_{D+1}+\nonumber\\
iGF_{\mu\nu}\xi^\mu\xi^\nu\xi_{D+1} \approx 0.
\end{equation}

Consider now a theory with Dirac brackets
(\ref{PrelimDB}) and a hamiltonian
\begin{equation}
\label{PrelHam}
H = \frac{i\chi}{2}\Phi_{D+1},
\end{equation}
where $\chi$ is the Grassmann odd Lagrange
 multiplier to $\Phi_{D+1}$. The consistency
  condition will lead to a new constraint
\begin{eqnarray}
\label{ConstraintsD2}
&& {\{\Phi_{D+1},\Phi_{D+1}\}}^*  =
i\Phi_{D+3} = i({\cal P}_\mu {\cal P}^\mu -
ig\frac{M}{2}F_{\mu\nu}\xi^\mu\xi^\nu+ \nonumber\\
&& +4iGF_{\mu\nu}{\cal P}^\mu\xi^\nu\xi_{D+1}+
G^2{(F_{\mu\nu}\xi^\mu\xi^\nu)}^2 - m^2) \approx 0,
\end{eqnarray}
where $g\frac{M}{2}=(g-2Gm)$,$M$ (as it will be
shown below) is the total magnetic moment of the
particle.
Now the extended hamiltonian is given by
\begin{equation}
\label{ExtHam}
H_{\rm ext} = \frac{i\chi}{2}\Phi_{D+1}+\frac{e}{2} \Phi_{D+3},
\end{equation}
where $e$ is Grassmann even Lagrange multiplier to
the consfraint $\Phi_{D+3}$.
The consistency equations imply no new constraints
since due to Jacoby identity
\begin{equation}
\label{JacobyId}
{\{\Phi_{D+3},\Phi_{D+1}\}}^*={\{{\{\Phi_{D+1},
\Phi_{D+1}\}}^*,\Phi_{D+1}\}}^*=0.
\end{equation}

The action of the relativistic spinning particle
with AMM in the external electromagnetic field
can be found by Legendre transformation
using (\ref{ExtHam}) and the constraints
(\ref{ConstraintsA}).In this way we find that
\begin{eqnarray}
\label{Action}
&& L = {\shalf}\int d\tau\Bigl[\frac{\left(\dot x^\mu\right)^2}
{e}+em^2-i\left(\xi_\mu\dot \xi^\mu-
\xi_{D+1}\dot \xi_{D+1}\right)
 +2g\dot x^\mu A_\mu+ig\frac{M}{2}e
 F_{\mu\nu}\xi^\mu\xi^\nu- \nonumber\\
&& -4iGF_{\mu\nu}\dot x^\mu\xi^\nu\xi_{D+1}-
 i\chi \bigl(\frac{\xi_\mu\dot x^\mu}{e}-m\xi_{D+1}
-iGF_{\mu\nu}\xi^\mu\xi^\nu\xi_{D+1}\bigr)-
eG^2(F_{\mu\nu}\xi^\mu\xi^\nu)^2
\Bigr],
\end{eqnarray}
where  the overdote denotes the differentiation
over $\tau$ along the  trajectory; the derivatives
 over Grassmann variables are left.
 As we'll see, this action leads to a
Bargmann-Michel-Telegdi equation of motion for the spin
$\shalf$ particle .

\section{The Bargmann-Michel-Telegdi equation}
\indent

Now we will use the hamiltonian (\ref{ExtHam}) to deduce
the equation of motion of the spin of the particle
with AMM in the external electromagnetic field.
To describe the evolution of spin
consider the Pauli-Lubanski vector in $D=2n$
dimensions \cite{GG1}
\begin{equation}
\label{PLVector}
W_\mu= \frac{(-i)^{\frac{D-2}{2}}}{(D-2)!}
\varepsilon_{\mu\nu\la_2\la_3\ldots \la_{D-1}}
{\cal P}^\nu \xi^{\la_2}\xi^{\la_3}\ldots \xi^{\la_{D-1}}.
\end{equation}
To obtain the covariant BMT equation we will use
(as it is customary) the gauge $\chi=0,e=\frac{1}{m}$
which  was used to deduce the BMT equation for
 spinning particle without AMM in $D=4$ dimensions
 \cite{BM},\cite{BCL},\cite{GT}. We have
\begin{equation}
\label{DerPLVec}
\dot W_\mu= \frac{(-i)^{\frac{D-2}{2}}}{(D-2)!}
\varepsilon_{\mu\nu\la_2\la_3\ldots \la_{D-1}}
\left(\dot {\cal P}^\nu\xi^{\la_2}+
(D-2){\cal P}^\nu\dot \xi^{\la_2}\right)
\xi^{\la_3}\cdots \xi^{\la_{D-1}}.
\end{equation}
Using  Eq.(\ref{ExtHam}) we find the expressions
for $\dot {\cal P}^\nu$ and $\dot \xi^{\la_2}$
\begin{eqnarray}
\label{DerMom}
&&\dot {\cal P}^\nu  =  {\{{\cal P}^\nu,H_{\rm ext}\}}^*=
\frac{1}{2m}\bigl[2g{\cal P}^\mu F^\nu{}_{\!\mu}+
ig\frac{M}{2}\pa^\nu F_{\mu\la}\xi^\mu\xi^\la
+4iG F^\nu {}_{\!\mu} F^\mu{}_{\!\la}
\xi^\la \xi_{D+1} - \nonumber\\
 && -4iG\pa^\nu F_{\mu\la}{\cal P}^\mu\xi^\la\xi_{D+1} -
2G^2\pa^\nu F_{\mu\la}\xi^\mu\xi^\la
F_{\rho\d}\xi^\rho\xi^\d\bigr],
\end{eqnarray}
\begin{equation}
\label{DerXi}
\dot \xi^{\la_2}=
\frac{1}{2m}\left[gMF^{\la_2}{}_{\!\mu}\xi^\mu
+4GF^{\la_2}{}_{\!\mu}{\cal P}^\mu\xi_{D+1}+
4iG^2F^{\la_2}{}_{\!\mu}\xi^\mu
F_{\sigma\d}\xi^\sigma\xi^\d\right].
\end{equation}
Also we have tensor relations
\begin{equation}
\label{Tensor1}
\varepsilon^{\mu\nu\la_2\la_3\cdots \la_{D-1}}
W_\mu=(-i)^\frac{D-2}{2}\bigl(
{\cal P}^\nu\xi^{\la_2}\xi^{\la_3}\cdots \xi^{\la_{D-1}}+
(\rm cyclic\; permutations)\bigr),
\end{equation}
\begin{eqnarray}
\label{Tensor2}
&&\varepsilon^{\mu\nu\la_2\la_3\ldots \la_{D-1}}
 W_\mu{\cal P}_\nu  =  (-i)^\frac{D-2}{2}\Bigl[-
{\cal P}^2\xi^{\la_2}\xi^{\la_3}\ldots \xi^{\la_{D-1}}+
({\cal P}_\sigma\xi^\sigma)\bigl(
{\cal P}^{\la_2}\xi^{\la_3}\ldots \xi^{\la_{D-1}}-  \nonumber\\
 && -{\cal P}^{\la_3}\xi^{\la_2}\ldots \xi^{\la_{D-1}}+
{\cal P}^{\la_4}\xi^{\la_2}\ldots \xi^{\la_{D-1}}-\ldots
-{\cal P}^{\la_{D-1}}\xi^{\la_2}
\ldots \xi^{\la_{D-1}}\bigr)\Bigr] = \nonumber\\
 &&= (-i)^{\frac{D-2}{2}}\Bigl[-
m^2\xi^{\la_2}\xi^{\la_3}\ldots \xi^{\la_{D-1}}+
m\xi_{D+1}\bigl({\cal P}^{\la_2}\xi^{\la_3}
\ldots \xi^{\la_{D-1}}-  \\
 && -{\cal P}^{\la_3} \xi^{\la_2}\ldots \xi^{\la_{D-1}}+
{\cal P}^{\la_4}\xi^{\la_2}\ldots \xi^{\la_{D-1}}-\ldots
-{\cal P}^{\la_{D-1}}\xi^{\la_2}\ldots
 \xi^{\la_{D-2}} \bigr) \Bigr]+
O(\xi^{(D)}),  \nonumber
\end{eqnarray}
where the summand $O(\xi^{(D)})$ contains terms
of the order $\xi^{(D)}$, which after
quantization contribute corrections of the order
$\hbar$ to the BMT equation. The last equation was
obtained using Eqs.(\ref{ConstraintD1}),
(\ref{ConstraintsD2}). Taking into account
Eqs.(\ref{DerMom})-(\ref{Tensor2}) we find the
equation of motion of $W_\mu$
\begin{equation}
\label{EPLVector}
\dot W_\mu=g\frac{M}{2m}F_{\mu\nu}W^\nu+
2G\frac{{\cal P}_\mu}{m}F_{\nu\la}W^\nu
\frac{{\cal P}_\la}{m}+O (\xi^{(D)}).
\end{equation}
Now since
\begin{equation}
\label{Velocity}
u^\nu=\dot x^\nu={\{x^\nu,H\}}^* =
\frac{{\cal P}_\nu}{m}+
2i\frac{G}{m}
F^{\nu}{}_{\!\mu}\xi^\mu\xi_{D+1},
\end{equation}
we find from (\ref{EPLVector})
\begin{equation}
\label{EBMT}
\dot W_\mu=g\frac{M}{2m}F_{\mu\nu}W^\nu+
2Gu_\mu F_{\nu\la}W^\nu
u_\la+O(\xi^{(D)}).
\end{equation}

This equation after quantization is a generalization
to $D=2n$ dimensions of the BMT equation for the motion
of the spin of the particle with AMM
in an arbitrary external electromagnetic
field. Note that the Eq.(\ref{EBMT}) is form invariant
in all dimensions. From (\ref{EBMT}) it is
clear that M is a total magnetic moment of the
particle.

\section{Foldy-Wouthuysen Transformation}
\indent
Now we introduce the pseudoclassical canonical
 Foldy - Wouthuysen transformation with a generator
of  the  infinitesimal  canonical transformations
\cite{BCLFW}
\begin{equation}
\label{Generator}
S_{\rm cl}=-2i\left({\cal P}_j
\xi_j\right)\xi_{D+1}\theta,
\end{equation}
where $\theta$
is  a  function  of  the  variables   of   the   theory
which  will  be  specified  later.  The  result  of   the
finite canonical  transformation  of   any   dynamical
quantity  $f$   is given by the expression \cite{SM}
\begin{equation}
\label{CanonTrans}
\tilde f
=\widetilde {{e^{S_{\rm {cl}}}f}} = f+{\{f,S_{\rm cl}\}}^*+
\frac{1}{2!}{\{\{f,S_{\rm cl}\},S_{\rm cl}\}}^*+ \ldots ,
\end{equation}
Applying Eq.(\ref{CanonTrans}) to a function  A  of  the
independent  variables $x^i,{\cal P}_i,\xi^i$.   and taking into
account the relations
\begin{eqnarray}
&& {\{A\xi_{D+1},S_{\rm
 {cl}}\}}^*  =  A(2\theta)({\cal P}_j\xi_j),\\
 && {\{A({\cal P}_j \xi_j),S_{\rm {cl}}\}}^*
  =  -(2\theta)\gamma A\xi_{D+1}-
  2i({\cal P}_j \xi_j)\xi_{D+1} A
  {\{({\cal P}_j \xi_j),\theta\}}^*
  +2i{\{A,({\cal P}_j\xi_j)\}^*
  ({\cal P}_i\xi_i)\xi_{D+1}\theta  ,\\
&& {\{A({\cal P}_j\xi_j)\xi_{D+1},S_{\rm {cl}}\}}^*  = 0 ,
\end{eqnarray}
where $\gamma=i{\{({\cal P}_i\xi_i),({\cal P}_j\xi_j)\}}^*
= {\cal P}_i^2+igF_{ij}\xi_i\xi_j$ ,
 we find for $\tilde A$ the expression
\begin{eqnarray}
&&\tilde A  = A -
\frac{i}{\sqrt\c}}{\{A,({\cal P}_j\xi_j)\}^*
\xi_{D+1}sin(2\theta\sqrt\c) + \frac{i}{\c}
{\{A,({\cal P}_j\xi_j)\}}^*({\cal P}_k\xi_k)
\left(cos(2\theta\sqrt\c)-1\right)- \nonumber\\
&& -\frac{1}{{(\sqrt\c)}^3}
{\{{\{A,({\cal P}_k\xi_k)\}}^*,({\cal P}_j\xi_j)\}}^*
({\cal P}_i\xi_i)\xi_{D+1}
\bigl(sin(2\theta\sqrt\c)-2\theta\sqrt\c\bigr)- \\
&& - \frac{2}{\c}
{\{A,({\cal P}_j\xi_j)\}}^*{\{\theta,({\cal P}_j\xi_j)\}}^*
({\cal P}_k\xi_k)\xi_{D+1}\left(cos(2\theta\sqrt\c)-1\right)-
2i{\{A,\theta\}}^*({\cal P}_k\xi_k)\xi_{D+1}.\nonumber
\end{eqnarray}
. If  we now specify the function $\theta$ taking
 $tg(2\theta\sqrt\c)=\frac{\sqrt\c}{\tilde m}$, and hence
$sin(2\theta\sqrt\c)=\frac{\sqrt\c}{\omega}$,
$cos(2\theta\sqrt\c)=\frac{\tilde m}{\omega}$,
 where $\omega =
\sqrt{{{\cal P}_i}^2+{\tilde m}^2+igF_{ij}\xi_i\xi_j}=
\sqrt{\c+{\tilde m}^2}$,
 ${\tilde m}= m-iGF_{ij}\xi_i\xi_j$,then
\begin{eqnarray}
\label{TransVar}
&& \tilde A  =  A - i{\{A,({\cal P}_j\xi_j)\}}^*
\frac{\xi_{D+1}(\omega+\tilde m) +
 ({\cal P}_j\xi_j)}{\omega(\omega+\tilde m)}
+\frac{i{\{A,\c\}}^*({\cal P}_j\xi_j)\xi_{D+1}}
{2\omega^2(\omega+\tilde m)}+  \nonumber\\
&&  +\frac{{\{A,({\cal P}_j\xi_j)\}}^*
{\{({\cal P}_k\xi_k),\tilde m\}}^*({\cal P}_i\xi_i)\xi_{D+1}}
{\omega^3(\omega+\tilde m)}
+\frac{i{\{A,\tilde m\}}^*({\cal P}_j\xi_j)\xi_{D+1}}
{\omega^2}.
\end{eqnarray}
This is the expression of the result of the
Foldy-Wouthuysen transformation that will be
used below to derive the variables which diagonalize
 the final Dirac brackets.

\section{Final Dirac Brackets and Quantization}
\indent

Following \cite{GG3} we write down the complete
set of constraints which includes the
constraints which  fix the remaining gauge degrees of
freedom (compare with \cite{GG3}) :
\begin{equation}
\label{ConstraintA}
\Phi_\mu  =
\pi_\mu-\frac{i}{2}\xi_\mu,\mu=0,1,2,\ldots,D-1;\quad \Phi_D =
\pi_{D+1}+\frac{i}{2}\xi_{D+1},\nonumber\\
\end{equation}
\begin{eqnarray}
\label{ConstraintB}
&&  \Phi_{D+3}  =
{\cal P}_\mu {\cal P}^\mu -
ig\frac{M}{2}F_{\mu\nu}\xi^\mu\xi^\nu +
4iGF_{\mu\nu}{\cal P}^\mu\xi^\nu\xi_{D+1} +
G^2{(F_{\mu\nu}\xi^\mu\xi^\nu)}^2 - m^2,
\Phi_{D+4}=x_O^\prime; \nonumber\\
&&\hspace {3.5cm} \Phi_{D+5}=\pi_e,\quad
\Phi_{D+6}=e+\frac{1} {\tilde
\omega},
\end{eqnarray}
\begin{equation}
\label{ConstraintC}
\Phi_{D+1}={\cal P}_\mu\xi^\mu-m\xi_{D+1}+
iGF_{\mu\nu}\xi^\mu\xi^\nu\xi_{D+1} ,\quad
\Phi_{D+2}=a\xi_O+b\xi_{D+1},
\end{equation}    We have  by   definition
\begin{equation}
\label{FinalBracket}
{\{\tilde A,\tilde B\}}_{D(\Phi)}=
{\{\tilde A,\tilde B\}}^{**}
-{\{\tilde A,\varphi_r\}}^{**}C^{-1}_{rr^\prime}
{\{\varphi_{r^\prime},\tilde B\}}^{**}.
\end{equation}
Here $\varphi_r=(\Phi_{D+1},\Phi_{D+2}),
{\{\ldots,\ldots\}}^{**}$   stands for the Dirac brackets
 for   a subset   of constraints
 (\ref{ConstraintA}),(\ref{ConstraintB}): $C^{-1}$ is
   the  inverse matrix of
\begin{equation}
C_{rr^\prime}={\{\varphi_r,\varphi_{r^\prime}\}}^{**}
\end{equation}
Now we'll   take   an   advantage of  the
special structure  of  the constraints (\ref{ConstraintB}):
one  of each  pair of constraints is   a    canonical
variable.  This  allows  to prove   that
\begin{equation}
\label{PrelimDiracBracket}
{\{F,G\}}^{**}={\{F,G\}}^*
\end{equation}
for any     dynamical    variables
  $F$ and $G$ (see   e.g.\cite{GTY2}).
 With account    of Eq. (\ref{PrelimDiracBracket} ) the
 formula  (\ref{FinalBracket} )    takes    the   form
\begin{equation}
\label{FinalDiracBracket}
{\{\tilde A,\tilde B\}}_{D(\Phi)}=
{\{\tilde A,\tilde B\}}^*
-{\{\tilde A,\varphi_r\}}^*C^{-1}_{rr^\prime}
{\{\varphi_{r^\prime},\tilde B\}}^*.
\end{equation}
The  matrix    $C$ is given by
\begin{equation}
\label{DiracMatrix}
C_{rr^\prime}={\{\varphi_r,\varphi_{r^\prime}\}}^{**}=
\left(\matrix{O& i\alpha \cr i\alpha & i(a^2-b^2)}\right),
\end{equation}
where $ \a= - a\kappa\omega+bm $.
Also it is easy to prove that
\begin{equation}
{\{\tilde A,\tilde B\}}^*=
\widetilde {{\{ A, B\}}^*}
\end{equation}
and verify by direct calculation that
\begin{equation}
\label{FWD1}
{\{\tilde A,\Phi_{D+1}\}}^*\vert_{\Phi=0} =0.
\end{equation}
Then taking into account that $C^{-1}_{(D+2)(D+2)}=0$
 we find that
\begin{equation}
\label{FinalDBFW}
{\{\tilde A,\tilde B\}}_{D(\Phi)}=
\widetilde {{\{ A, B\}}^*}.
\end{equation}
Introducing the variables
\begin{equation}
\label{PrimedVar}
A^\prime \equiv \tilde A \vert_{\Phi = 0}=
A+i{\{A,({\cal P}_i\xi_i)\}}^*
({\cal P}_j\xi_j)\frac{(b\k+a)}
 {\tilde \b(\omega +\tilde  m)},
\end{equation}
and making use of the  property   of   the  Dirac brackets
 which  states, that   the Dirac brackets of   the
 constraints with any  dynamical variable vanish, we derive
 from Eq.(\ref{FinalDBFW}),(\ref{PrimedVar}) that
 \begin{eqnarray}
 \label{PrimedDB}
&&{\{\tilde A,\tilde B\}}_{D(\Phi)}=
{\{A^\prime,B^\prime\}}_{D(\Phi)}=
\left({\{A,B\}}^*\right)^\prime=\nonumber\\
&&={\{A,B\}}^*+
i{\{{\{A,B\}}^*,({\cal P}_i\xi_i)\}}^*
({\cal P}_j\xi_j)\frac{(b\k+a)}
 {\tilde \b(\omega +\tilde m)}.
\end{eqnarray}
If now  we take  for
$A,B$ the variables $x_i,{\cal P}_j ,\xi_k$,
 then on account of (\ref{PrimedVar}) we find from
\begin{eqnarray}
\label{PrimedIndVar}
&& x^\prime_i=x_i-i\xi_i ({\cal P}_j\xi_j)
 \frac{(b\k + a)}
 {\tilde \b(\omega + \tilde m)} \equiv q_i,  \nonumber\\
&& {\cal P}_i^\prime={\cal P}_i +
igF_{ij}\xi_j ({\cal P}_k\xi_k)\frac{(b\k+a)}
{\tilde \b(\omega + \tilde m)}\equiv\pi_i,\\
&&\xi_i^\prime = \xi_i + {\cal P}_i({\cal P}_j\xi_j)
\frac{(b\k+a)}
 {\tilde \b(\omega+\tilde m)} \equiv \psi_i, \nonumber
\end{eqnarray}
which for $D=4$
coincide with canonical variables found in \cite{GG3}
 after the diagonalization of Dirac brackets.
{}From (\ref{PrimedDB}) we find
\begin{eqnarray}
\label{CanonicalVar}
&&  {\{\psi_i,\psi_j\}}_{D(\Phi)}=
{\{\xi_i,\xi_j\}}^*=-i\d_{ij},\quad
{\{q_i,\pi_j\}}_{D(\Phi)}=
{\{x_i,{\cal P}_j\}}^*=-i\d_{ij},   \nonumber\\
&& {\{q_i,\psi_j\}}_{D(\Phi)}=
{\{x_i,\xi_j\}}^*=0,\quad {\{q_i,q_j\}}_{D(\Phi)}=
{\{x_i,x_j\}}^*=0,   \\
&&  {\{\pi_i,\pi_j\}}_{D(\Phi)}
=gF_{ij}(x)-ig\pa _kF_{ij}\xi^k
({\cal P}_m\xi_m)\frac{(b\k+a)}
{\tilde \b(\omega+\tilde m)}=gF_{ij}(q) . \nonumber
\end{eqnarray}
These relations prove that  the variables
 $\tilde x_i, \tilde {\cal P}_j,\tilde \xi_k$   ,
     are  Newton-Wigner  variables and
 the variables $q_i, \pi_j , \psi_k$ have canonical
Dirac brackets and hence it is  convenient to quantize
the theory in  terms
 of   these variables as  it was done for  $ D=4$.
 Expressions of  initial variables   in terms of
 canonical   ones are given by the  following
\begin{eqnarray}
\label{InitialVar}
&&  x_i  = q_i-i\psi_i(\pi_j\psi_j) \frac{(a\k + b)}
{\tilde \a(\Omega + \tilde m)},  \nonumber\\
&& {\cal P}_i=\pi_i + igF_{ij}\psi_j
(\pi_k\psi_k)\frac{(a\k+b)}
{\tilde \a(\Omega + \tilde m)},  \\
&& \xi_i = \psi_i + \pi_i (\pi_j\psi_j)\frac{(a\k+b)}
{\tilde \a(\Omega+\tilde m)},  \nonumber
\end{eqnarray}
where  $\tilde \a=-a\k\Omega+b\tilde m,\Omega=
\sqrt{\pi_i^2+{\tilde m}^2+igF_{ij}\psi_i\psi_j}$.
  Using (\ref{InitialVar}) and the relations
\begin{equation}
\xi_{D+1} =
-\frac{a(\pi_j\psi_j)}{\tilde \a},\quad
\xi_O=\frac{b(\pi_j\psi_j)}{\tilde \a},
\end{equation}
 one can  deduce the  expression for    the
  physical hamiltonian   of the spinning
    particle with AMM in  the external
    electromagnetic   field  in $D=2n$  in terms  of
canonical         variables
\begin{equation}
\label{Hamiltonian}
H_{\rm phys}=\Omega-g\k A_O
-ig\k\frac {F_{Ok}\psi_k(\pi_j\psi_j)}
{\Omega(\Omega+\tilde m)}+
\frac{2iG\k}{\Omega}
\left(F_{Ok}\psi_k+\frac{F_{ik}\pi_i\psi_k}
{(\Omega+\tilde m)}\right)(\pi_j\psi_j).
\end{equation}
Quantization of  the theory  is
similar  to that in $D=4$.  It is   worth mentioning  that
  to quantize  the  theory   by  the   Berezin,  Marinov
prescription [4]  one must  expand $\Omega,\tilde m$,
 which enters
e.g.   (\ref{Hamiltonian} ),   in powers of
 $F_{ij}\psi_i\psi_j$, the  expansion terminating  in the
order $\frac{D-2}{2}$. If however  we are  interested in
the   terms in   the  hamiltonian which after  quantization
  will  be of  the order  of  $\hbar$,   then  taking into
 account  that after  quantization
  $\psi_i \Rightarrow {\sqrt\frac{\hbar}{2}}\sigma_i$,
 we find    for   the quantum Hamiltonian the expression
\begin{eqnarray}
\label{PhysHam}
&&  \hat H_{phys}
 {\rightarrow \atop \leftarrow}
 \tilde \Omega -g\k A_O -ig\k\hbar
\frac{F_{Ok}\pi_j(\sigma_k\sigma_j-\sigma_j\sigma_k)}
{4\tilde \Omega( \tilde \Omega+m)}+  \nonumber\\
&&  +ig\hbar\frac{F_{ij}\sigma_i\sigma_j}{4\tilde \Omega}+
\frac{iG\k}{2\tilde \Omega}(\sigma_k\sigma_j-\sigma_j\sigma_k)
\Bigl(F_{Ok}\pi_j+\frac{F_{ij}\pi_i\pi_k}
{(\tilde \Omega+m)}\Bigr),
\end{eqnarray}
where $\tilde \Omega=\sqrt{\pi_i^2+m^2}$,
$\rightarrow \atop \leftarrow$
denotes the Weyl correspondence
 between operators and their symbols.

The expression for the hamiltonian
(\ref{PhysHam}) in $D=4$ dimensions in the
first order of g coincides with the hamiltonian
 found in \cite{SG}. Note however that
 the hamiltonian (\ref{PhysHam}) is correct
 in all orders of g and was obtained without
 restrictions on potentials.

\vspace{5mm}
\noindent{\em Acknowledgment}

The authors wish to thank I.V.Tyutin
for useful discussions.

 This research was partly supported by the grant
  YPI-1993 of the "Bundesminister f\"ur Forschung und
Technologie", Federal Republic of Germany.

\end{document}